# Amplitudes of stellar oscillations: the implications for asteroseismology

**H. Kjeldsen*** and **T. R. Bedding**

European Southern Observatory, Karl-Schwarzschild-Str. 2, D-85748 Garching bei München, Germany

**Abstract.** There are no good predictions for the amplitudes expected from solar-like oscillations in other stars. In the absence of a definitive model for convection, which is thought to be the mechanism that excites these oscillations, the amplitudes for both velocity and luminosity measurements must be estimated by scaling from the Sun. In the case of luminosity measurements, even this is difficult because of disagreement over the solar amplitude.

This last point has lead us to investigate whether the luminosity amplitude of oscillations ($\delta L/L$) can be derived from the velocity amplitude ($v_{\rm osc}$). Using linear theory and observational data, we show that $p$-mode oscillations in a large sample of pulsating stars satisfy $(\delta L/L)_{\rm bol} \propto v_{\rm osc}/T_{\rm eff}$. Using this relationship, together with the best estimate of $v_{{\rm osc},\odot} = (23.4 \pm 1.4)\,{\rm cm\,s}^{-1}$, we estimate the luminosity amplitude of solar oscillations at 550 nm to be $(\delta L/L) = (4.7 \pm 0.3)$ ppm.

Next we discuss how to scale the amplitude of solar-like (i.e., convectively-powered) oscillations from the Sun to other stars. The only predictions come from model calculations by Christensen-Dalsgaard & Frandsen (1983, Sol. Phys. 82, 469). However, their grid of stellar models is not dense enough to allow amplitude predictions for an arbitrary star. Nevertheless, although convective theory is complicated, we might expect that the general properties of convection – including oscillation amplitudes – should change smoothly through the colour-magnitude diagram. Indeed, we find that the velocity amplitudes predicted by the model calculations are well fitted by the relation $v_{\rm osc} \propto L/M$.

These two relations allow us to predict both the velocity and luminosity amplitudes of solar-like oscillations in any given star. We compare these predictions with published observations and evaluate claims for detections that have appeared in the literature. We argue that there is not yet good evidence for solar-like oscillations in any star except the Sun. For solar-type stars (e.g., $\alpha$ Cen A and $\beta$ Hyi), observations have not yet reached sufficient sensitivity to detect the amplitudes we predict. For some F-type stars, namely Procyon and several members of M67, detection sensitivities 30–40% below the predicted amplitudes have been achieved. We conclude that these stars must oscillate with amplitudes less than has generally been assumed.

**Key words:** Sun: oscillations – Stars: individual: $\alpha$ Cen – Stars: individual: Procyon – Stars: oscillations – Cepheids – $\delta$ Scu

## 1. Introduction

Asteroseismology of solar-like stars has so far produced disappointing results. Despite repeated attempts and ever-increasing sensitivity, there have been no unambiguous detections of solar-like oscillations on any star except the Sun. The most recent effort involved seven 4 m-class telescopes and reached unprecedented detection thresholds on twelve stars in M67, but still gave only suggestive evidence for oscillations (Gilliland et al. 1993).

A crucial problem with interpreting the observations is our poor knowledge of the amplitude of the expected oscillations. When designing experiments, choosing targets and discussing results, it is clearly important to have a good prediction for the amplitude of the signal one is trying to detect. Addressing this issue is the main aim of this paper.

Since the Sun is the only star for which we have a positive detection, it is the obvious starting point from which to estimate amplitudes for other stars. Solar oscillations are thought to be excited by turbulent convection near the surface (see Murray 1993 for a recent review). Stars with effective temperatures below $\sim 7000$ K have a convective layer in their outer parts (Gray & Nagel 1989) and so might be expected to undergo solar-like oscillations. Observations must be made in light integrated over the stellar disk, which means that only the low-degree modes will be observable. Fortunately, these modes are particularly interesting because they penetrate deeply into the stellar interior. Two methods have developed to search for oscillations. The first is to look for periodic brightness fluctuations using photometry (*luminosity* measurements), while the other involves searching for periodic Doppler shifts of spectral features (*velocity* measurements). In both cases, the aim is to measure the frequencies of the oscillation modes, since these would tell us a great deal about the physical conditions inside the star.

One can estimate amplitudes for stellar oscillations by scaling from the solar case. This has been done by Christensen-Dalsgaard & Frandsen (1983; hereafter C-DF), who have computed theoretical models for stars in different parts of the colour–magnitude diagram. For the case of *luminosity* amplitudes, however, predicting values for other stars is impeded by the fact that we do not yet have a good value for the Sun. It has turned out to be very difficult to measure the luminosity amplitude of low-degree solar oscillations and the current value is uncertain by more than a factor of two. This directly affects the predictions for other stars, which is unfortunate because differential photometry – as carried out by Gilliland et al. (1993) on M67 – is one of the most promising methods for detecting stellar oscillations. The *velocity* amplitude of solar oscillations, on the other hand, is quite well known. This has prompted us to investigate whether there is a simple relationship between velocity and luminosity amplitudes of stellar oscillations. In Sect. 2 we derive such a relationship and use it to make an improved estimate for the luminosity amplitude of oscillations in the Sun.

Even with accurate solar amplitudes, scaling to other stars is not straightforward. Model calculations are complicated and the only ones available are those of C-DF. Their predicted amplitudes generally increase with stellar temperature and luminosity, but the grid of stellar models is not dense enough to allow amplitude predictions for an arbitrary star. In Sect. 3 we discuss how to scale the *velocity* amplitude of solar oscillations to other stars, using only the fundamental stellar parameters. Combining this with the results of Sect. 2 also us allows to predict *luminosity* amplitudes. In Sect. 4, we review the other property of oscillations that can be predicted by scaling from the Sun, namely the frequency spectrum of the modes. Finally, in Sect. 5 we compare the predictions of both amplitudes and frequencies with published observations of solar-like oscillations in other stars (both upper limits and claimed detections).

There are several important issues that must be considered when measuring the amplitudes of oscillations. These include the calibration of power spectra, the effect of noise on amplitude estimates and the effect of finite mode lifetimes. We discuss these points in the Appendix, so as not to interrupt the flow of the paper.

## 2. Relating velocity and luminosity amplitudes

### 2.1. Linear adiabatic theory

In this section we seek to derive a relationship between velocity and luminosity amplitudes of stellar oscillations. Note that we define an oscillation to be 'solar-like' if it is excited by convection in the outer part of the star. Although the Sun is the only star for which we have reliable measurements of solar-like oscillations, there are many classes of stars which show other types of oscillatory behaviour. Oscillations in these stars are not excited by convection, but by the so-called $\kappa$ mechanism (see, e.g., Cox 1985). This is a feedback process and the resulting amplitudes are much larger than those of convectively-driven oscillations, making them easier to observe. We note in passing that the difficulty with interpreting the frequency spectra of these large-amplitude oscillators lies in deciding which modes are being excited (Matthews 1993). This is less of a problem for solar-like oscillations because one expects all the modes in a large frequency range to be excited.

Stars that undergo classical (i.e., non-solar-like) oscillations include $\beta$ Cephei stars, $\delta$ Scuti stars, classical Cepheids (also called $\delta$ Cephei stars) and RR Lyrae stars. Although the excitation mechanisms and amplitudes of these oscillations are very different from the solar case, the physics is not. In all cases, we are observing acoustic $p$-mode oscillations, in which the dominant restoring force is pressure. Other types of oscillations, such as the $g$-mode (gravity) waves found in white dwarfs, are not considered in this paper. Acoustic $p$-mode oscillations can be described to first order by simple linear theory (Cox 1980). The following analysis should be applicable to all these classes of oscillating stars, provided that the amplitude of the perturbations is not too large.

Let $(\delta L/L)_{\rm bol}$ be the fractional variation in the bolometric luminosity of the star. This luminosity variation is due almost entirely to changes in the temperature (the change in radius is negligible), so

$$(\delta L/L)_{\rm bol} \propto \frac{\delta T}{T}. \tag{1}$$

Under the assumption that the oscillations are adiabatic, we have

$$\frac{\delta T}{T} \propto \frac{\delta \rho}{\rho}.$$

In other words, the luminosity variations scale directly with the relative compression of the atmosphere.

To first order, the density compression for an adiabatic sound wave in a medium with sound speed $c_{\rm s}$ is (Landau & Lifshitz 1959)

$$\frac{\delta \rho}{\rho} = \frac{v}{c_{\rm s}},$$

where $v$ is the fluid velocity (i.e., the physical velocity of the gas). For the case of stellar oscillations, the fluid velocity determines the velocity amplitude that we observe, so that we have $v_{\rm osc} \propto v$.

The only thing that remains is to estimate the adiabatic sound speed. Using

$$c_{\rm s}^2 = \left[ \frac{\partial \ln P}{\partial \ln \rho} \right]_{\rm ad} \cdot \frac{P}{\rho}$$

and the ideal gas law $P \propto \rho T$ gives

$$c_s^2 \propto T. \tag{2}$$

Here, we have made the assumption that the adiabatic gradient $\left[\partial \ln P / \partial \ln \rho\right]_{ad}$ is constant. This is true for a fully ionized gas (where $\left[\partial \ln P / \partial \ln \rho\right]_{ad} = 5/3$), and also holds quite well in the outer parts of cool stars (Kippenhahn & Weigert 1990).

In Eq.(2), $T$ is the mean local temperature. We observe the oscillations in the photosphere, where $T$ is close to the effective temperature and only changes slowly as function of pressure and optical depth. Therefore, in going from one star to another it should be valid to set $T \propto T_{eff}$.

Putting all this together, we finally have

$$(\delta L/L)_{bol} \propto \frac{v_{osc}}{\sqrt{T_{eff}}}. \tag{3}$$

This is the relation we seek: it expresses the luminosity amplitude of a stellar oscillation in terms of the velocity amplitude. However, we have used a very simple adiabatic model of the stellar atmosphere and a linear theory which assumes that the oscillations are small in amplitude. While this last assumption is certainly true for solar-like oscillations, it is more questionable for classical oscillating stars. In the next section we compare our result with observational data in an effort to verify the relation and derive the constant of proportionality.

### 2.2. Calibrating the relationship

To test Eq.(3), we have compiled observational data for 90 oscillating stars of the following types: 12 $\beta$ Cephei stars (Lesh 1982), 5 $\delta$ Scuti stars (Danziger & Kuhi 1966; Breger et al. 1976; Campos & Smith 1980; Smith 1982), 43 $\delta$ Cephei stars (Moffett & Barnes 1980, 1984, 1985; Barnes et al. 1988b; Wilson et al. 1989) and 30 RR Lyrae stars (Manduca et al. 1981; Barnes et al. 1988a,b; Liu & Janes 1990a,b; Clementini et al. 1994).

In order to make the comparison, we must allow for the fact that the luminosity amplitude of an oscillation depends on the wavelength at which it is observed (larger amplitudes are observed at shorter wavelengths). We therefore need to convert each observed luminosity amplitude $(\delta L/L)_\lambda$ into a bolometric amplitude $(\delta L/L)_{bol}$.

To do this, we model the stellar spectrum as a black body and use our assumption that the luminosity variation is due entirely to changes in temperature (Eq. 1). It is then straightforward to show that

$$\begin{aligned}(\delta L/L)_{bol} &= (\delta L/L)_\lambda \, \frac{\lambda}{\lambda_{bol}} \\ &\propto (\delta L/L)_\lambda \, \lambda \, T_{eff}.\end{aligned} \tag{4}$$

Here

$$\lambda_{bol} = \frac{623\,\text{nm}}{T_{eff}/5777\,\text{K}}$$

is the wavelength at which the observed luminosity amplitude is exactly equal to $(\delta L/L)_{bol}$. Multi-wavelength observations confirm that the black-body model is a good approximation for both classical oscillations (Clementini et al. 1994) and for oscillations in the Sun (Schrijver et al. 1991; Toutain & Gouttebroze 1993). Equation (4) is accurate to a few percent, provided that $\lambda$ is not too different from $\lambda_{bol}$ (within $\pm 40\%$). All 90 data points in our sample satisfy this condition.

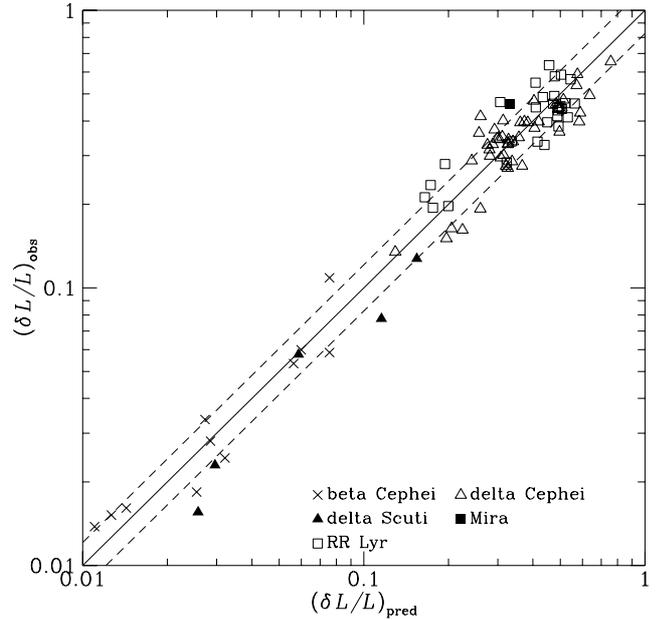

**Fig. 1.** Observed luminosity amplitudes of a sample of oscillating stars, plotted against values predicted from their velocity amplitudes. The dashed lines show the $1\sigma$ scatter of 21%.

Combining Eqs.(3) and (4) gives

$$(\delta L/L)_\lambda \propto \frac{v_{osc}}{\lambda} T_{eff}^{-1.5}.$$

We have compared this relation with the observed luminosity and velocity amplitudes for the 90 stars in our sample. The agreement is good, but can be improved if we allow the exponent to differ from $-1.5$. A least-squares fit gives $-2.0$ as the best value. As noted above, we have used a very simple adiabatic description of the stellar atmosphere and it is therefore not surprising that we need to include a correction that depends on temperature.

Fitting this revised equation to the data gives the following calibrated relations:

$$\begin{aligned}(\delta L/L)_\lambda &= \frac{v_{osc}/\text{m s}^{-1}}{(\lambda/550\,\text{nm})\,(T_{eff}/5777\,\text{K})^2}\,20.1\,\text{ppm} \tag{5} \\ (\delta L/L)_{bol} &= \frac{v_{osc}/\text{m s}^{-1}}{T_{eff}/5777\,\text{K}}\,17.7\,\text{ppm},\end{aligned}$$

where ppm denotes parts-per-million (note that 1 ppm equals $1.086\,\mu\text{mag}$).

### 2.3. Discussion

Figure 1 shows the relation for the stars in our sample: we have used Eq.(5) to predict $(\delta L/L)_\lambda$ for each star and we compare these values with the observations. For comparison we also show Mira (Hoffmeister et al. 1985) which shows remarkably good agreement with the relation. It appears that Eq.(5) remains a good approximation, even for oscillations which are far from being either linear or adiabatic. This seems to indicate that other physical processes are involved, and deserves further investigation. Of course, the observed correlation is still useful for our purpose, even though its validity for the extreme cases is unexplained.

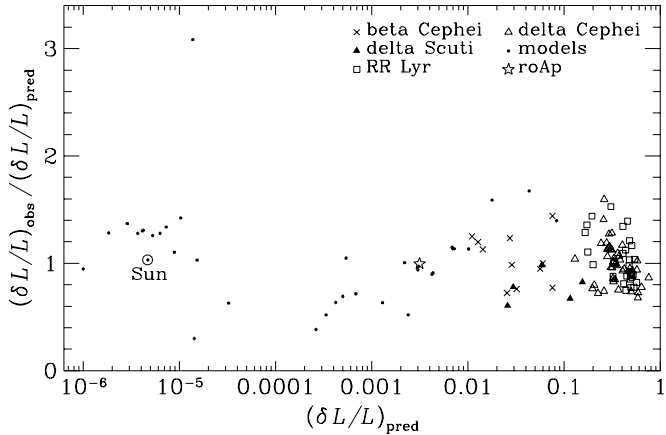

**Fig. 2.** Observed divided by predicted luminosity amplitudes for our sample of 90 stars, one roAp star, 35 model stars from C-DF and the Sun (Schrijver et al. 1991).

The 1-$\sigma$ scatter in Fig. 1 (excluding Mira) is 21%. Part of this can be attributed to measurement errors on the data points. In particular, $T_{eff}$ is only known to 5–10% for most of the stars and, combined with the uncertainties in velocity amplitudes, this is enough to explain most of the scatter. In this case, the accuracy of the calibrated relationship is $21\%/\sqrt{90} = 2.2\%$. We will adopt this value as the uncertainty in Eq.(5).

Can we apply Eq.(5) to solar-like oscillations? These have much smaller amplitudes, so we expect the linear theory to be a very good approximation. Observations of the Sun support this argument: the power spectrum of solar oscillations has a regular series of peaks modulated by a broad envelope (see Fig. 4 below) and, as we expect, the envelopes for luminosity and velocity measurements are very similar. To quantify this, Schrijver et al. (1991) have used simultaneous measurements of the solar luminosity amplitude (at 500 nm using the IPHIR instrument on the PHOBOS 2 spacecraft) and the velocity amplitude (from ground-based observations) to calculate the ratio between the two. Note that this was a differential calculation, since they were only concerned with ratios and not absolute vales. They find the ratio to be the same for all modes, having an average value of $(22.8 \pm 1.5)$ ppm/m s$^{-1}$. From Eq.(5) (assuming 2.2% accuracy), we would have predicted $(\delta L/L)_{500}/v_{osc} = 22.1 \pm 0.5$ ppm/m s$^{-1}$. The excellent agreement gives very strong confirmation that our relation, which was calibrated without using any data from the Sun, can be applied to solar oscillations.

For stars other than the Sun, the only 'data' on solar-like oscillation amplitudes come from the models of C-DF. Their main aim was to calculate velocity amplitudes using convective theory, as we discuss in Sect. 3 below. However, they also used a linear calculation to estimate luminosity amplitudes from their velocity amplitudes. We have repeated this calculation of $(\delta L/L)$ from $v_{osc}$ by using our relation.

Figure 2 shows the ratio between 'observed' and 'predicted' luminosity amplitudes. For the models, 'observed' values refer to the linear calculation by C-DF and 'predicted' refers to our calculation. The agreement is reasonable but the scatter is quite high. For the models with lowest amplitudes, including the model for the Sun at age 4.75 Gyr, there is a tendency for the calculations of $(\delta L/L)$ by C-DF to be higher than Eq.(5). This contrasts with the good agreement we found with the Schrijver et al. measurement of the Sun (see Fig. 2). We believe that our results give a more correct description than the calculation by C-DF, which was based on a simple atmospheric model.

### 2.3.1. Rapidly oscillating Ap stars

Some peculiar Ap stars exhibit oscillations with periods from $\sim 4$ to 20 minutes (see Kurtz 1990 and Matthews 1991 for recent reviews). The driving mechanism is probably related to rotation and magnetic fields, but is not well understood. Nevertheless, the oscillations are thought to be $p$-mode and so Eq.(5) should apply. Typical luminosity amplitudes are a few milli-magnitudes (mmag), making these oscillations intermediate between classical pulsations and solar-type oscillations and thus providing a useful test for our relation.

Making the test requires simultaneous luminosity and velocity measurements, since the oscillations vary greatly in amplitude from day to day. Unfortunately, velocity oscillations have only been detected on two stars, and in only one case (HR 1217) were simultaneous photometric measurements also obtained (Matthews et al. 1988). Measurements on one night gave $(\delta L/L) = 3100$ ppm (6.8 mmag peak-to-peak) in Johnson $B$ (440 nm) and $v_{osc} = (200 \pm 25)$ m s$^{-1}$. We estimate the effective temperature of the star to be $(7300 \pm 300)$ K, based on its Johnson and Strömgren colours. This is in agreement with the value of 7400 K given by Kurtz & Martinez (1993). From the measured $v_{osc}$ and $T_{eff}$, we use Eq.(5) to predict a luminosity amplitude of $(\delta L/L)_{440} = (3150 \pm 450)$ ppm, in good agreement with the observed value (see Fig. 2).

On a second night the luminosity amplitude was much lower (1.8 mmag peak-to-peak), from which we expect the velocity amplitude to be $\sim 50$ m s$^{-1}$. The observations gave an upper limit of 65 m s$^{-1}$, consistent with the predicted value, and we note that there is some indication of velocity oscillations at the expected level (Fig. 3b of Matthews et al. 1988).

We conclude that, despite the peculiar nature of Ap oscillations, the velocity-to-luminosity amplitude ratio for HR 1217 is well described by Eq.(5).

**Table 1.** Velocity amplitudes of oscillations in the Sun

| Reference | $v_{osc,\odot}$ (cm s$^{-1}$) |
|---|---|
| Grec et al. (1983) | 23.0 |
| Isaak et al. (1989) | 21.0 |
| Jiménez et al. (1990) | 24.0 |
| Libbrecht & Woodard (1991) | 25.5 |

**Table 2.** Luminosity amplitudes of oscillations in the Sun

| Reference | $(\delta L/L)_{550}$ (ppm) |
|---|---|
| Woodard & Hudson (1983a,b) | 4.7 |
| Jiménez et al. (1990) | 3.7–6.5 |
| Toutain & Fröhlich (1992) | 3.6 |
| this paper | 4.7 ± 0.3 |

### 2.4. The solar amplitude

As discussed in the Introduction, there is disagreement over the luminosity amplitude of low-order solar oscillations. The velocity amplitude is better known, so we now use Eq.(5) to estimate $(\delta L/L)$ from $v_{osc}$ for the Sun.

When considering published measurements of $v_{osc,\odot}$, we must keep in mind that the amplitudes of individual modes can vary by a factor of $\sim 2$ over time scales of a few days. On the other hand, there are many different modes varying independently and the maximum amplitude, averaged for several days over the strongest few modes, stays roughly constant. Since we are concerned with detecting oscillations on other stars, it is this average maximum amplitude that is of interest to us. Table 1 lists the measurements and Appendix A.4 gives details on how we arrived at these values from the published data. The mean value is $v_{osc,\odot} = (23.4 \pm 1.4)$ cm s$^{-1}$. Putting this into Eq.(5) gives a luminosity amplitude at 550 nm of $(4.7 \pm 0.3)$ ppm.

Three groups have observed solar oscillations in luminosity. Taking the published measurements directly gives peak values of $(\delta L/L)$ ranging from 2 to 11 ppm. However, some important corrections are required to convert these values into true oscillation amplitudes, as described in Appendix A.4. The values after correction are summarized in Table 2: the first and third are from spacecraft data and the second comes from ground-based observations. In fact, after the corrections have been made, the disagreement between the measurements has largely disappeared. We adopt our calculated value as the most accurate.

With our present poor understanding of the process that excites solar oscillations, an improved estimate of their amplitude will not aid our understanding of the Sun very much. The purpose of this paper is to estimate amplitudes of solar-like oscillations in other stars and, for this, the new result is very useful. However, in order to make these estimates, we must also develop a way of scaling the solar amplitude to other stars.

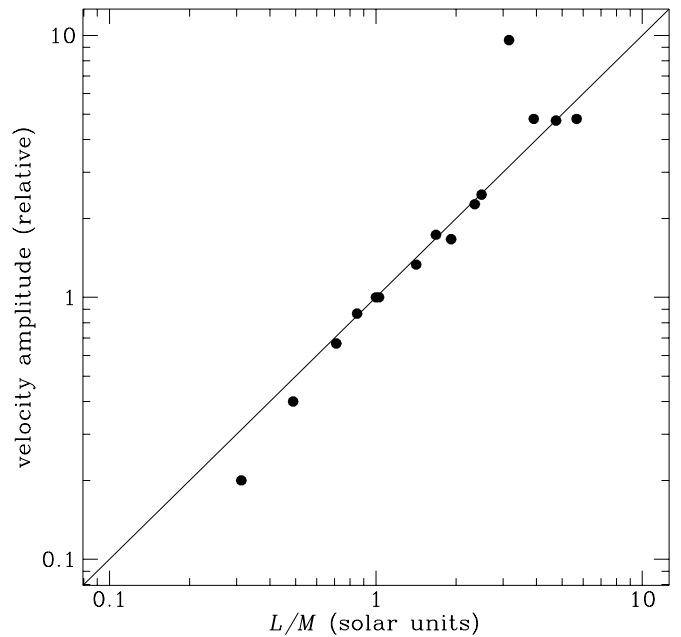

**Fig. 3.** Velocity amplitude versus light-to-mass ratio for solar-like oscillations in fifteen stellar models calculated by C-DF. The amplitudes are normalized relative to their model of the Sun.

## 3. Scaling velocity amplitudes to other stars

### 3.1. Models of Christensen-Dalsgaard & Frandsen (1983)

We wish to estimate the velocity amplitude $v_{osc}$ of solar-like oscillations by scaling from the Sun. The solar oscillations are thought to be excited in the outer part of the convection zone, very near the surface, where the motion of convective elements triggers pressure waves in the star (see Murray 1993 for a recent review). The damping mechanism is probably also related to convection via viscosity and scattering processes.

The only predictions for amplitudes of solar-like oscillations come from model calculations by C-DF. They calculated amplitudes for stars in different parts of the colour–magnitude diagram, but the grid of stellar models is not dense enough to allow amplitude predictions for an arbitrary star. However, we expect that the basic properties of convection change smoothly through the colour-magnitude diagram. We therefore postulate the existence of a scaling relationship with which one could predict oscillation amplitudes as a function of the fundamental stellar parameters.

As shown in Fig. 3, we find that the velocity amplitudes calculated by C-DF for main sequence stars and subgiants are well fitted by the relation

$$v_{osc} \propto L/M. \qquad (6)$$

That is, the velocity amplitude appears to scale directly with the light-to-mass ratio of the star. The agreement is fairly good, but there are several discrepant points. We believe these are due to the details of the C-DF models, as follows.

The oscillation spectra predicted by C-DF have envelopes (i.e., amplitude as a function of oscillation frequency) that differ substantially from that observed on the Sun. This is presumably because their treatment of excitation and damping is incomplete. For the low-mass stars ($0.8$–$1.3\,M_\odot$), their oscillation spectra rise monotonically until a cutoff frequency, where they drop to zero. The peak amplitude of each spectrum therefore depends critically on this cutoff frequency, which they calculate as the inverse of the minimum timescale for convective eddies. In particular, reducing the cutoff frequency means underestimating the peak amplitude. For the lowest mass stars in their models, the cutoff frequency they derive actually decreases with decreasing mass. This is the opposite of what is expected from using the acoustic cutoff frequency (see Sect. 4.2), and can explain why the bottom two stars in Fig. 3 fall below the relation.

For stars with higher mass ($1.4$–$1.8\,M_\odot$), their oscillation spectra have broad flat envelopes and so the peak amplitude is not sensitive to the value of the cutoff frequency. Their results for these stars are in good agreement with Eq.(6), except for one case. The discrepant model ($1.5\,M_\odot$) has an oscillation spectrum with a very different envelope: the peak is narrower and at a lower frequency than the other models. This model corresponds to the point falling far above the relation in the figure.

Apart from these few discrepant points, the C-DF models are well fitted by Eq.(6). Until new models are produced that include better descriptions of the excitation and damping mechanisms, we conclude that Eq.(6) is the best we can do for predicting the amplitudes of solar-like oscillations.

### 3.2. Discussion

What is Eq.(6) telling us about the physics of the oscillations? We can use $g \propto M/R^2$ and $L \propto R^2 T_{\mathrm{eff}}^4$ to write the relation as

$$v_{\mathrm{osc}} \propto T_{\mathrm{eff}}^4/g,$$

where $g$ is the surface gravity. Since convection is the dominant mechanism for energy transport in the convection zone, the radiative surface flux ($\sigma T_{\mathrm{eff}}^4$) must equal the convective flux $F_{\mathrm{con}}$, so

$$v_{\mathrm{osc}} \propto F_{\mathrm{con}}/g$$
$$\propto F_{\mathrm{con}} H_{\mathrm{P}}/T.$$

Here $H_{\mathrm{P}} \propto T/g$ is the pressure scale height (Kippenhahn & Weigert 1990). We see that, provided Eq.(6) holds, the velocity amplitude of oscillations is determined by the convective flux, the scale height and the temperature.

### 3.3. Predicting oscillation amplitudes

To summarize, our prediction for velocity oscillation amplitudes is:

$$v_{\mathrm{osc}} = \frac{L/L_\odot}{M/M_\odot}\,(23.4 \pm 1.4)\,\mathrm{cm\,s^{-1}}, \qquad (7)$$

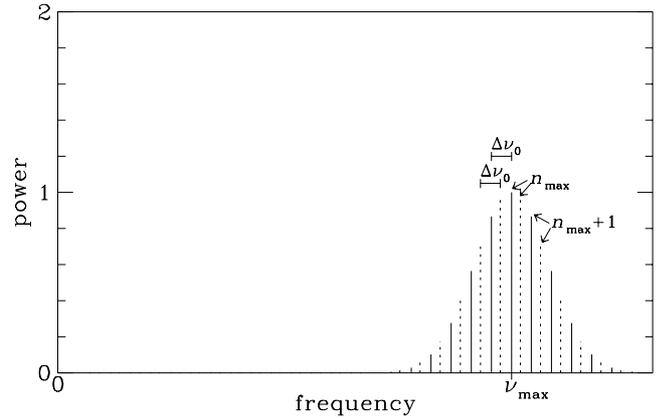

**Fig. 4.** Schematic diagram of the power spectrum of solar oscillations. Each peak corresponds to an oscillation mode: solid peaks are $l = 0$ modes and dashed peaks are $l = 1$.

where we have used Eq.(6) and the measured value of $v_{\mathrm{osc},\odot}$ (Sect. 2.4). We can then use Eq.(5) to predict the luminosity amplitude:

$$(\delta L/L)_\lambda = \frac{L/L_\odot\,(4.7 \pm 0.3)\,\mathrm{ppm}}{(\lambda/550\,\mathrm{nm})\,(T_{\mathrm{eff}}/5777\,\mathrm{K})^2(M/M_\odot)}. \qquad (8)$$

## 4. Oscillation frequencies

Before comparing the predictions given in Sect. 3.3 with published observations, we briefly review several other oscillation properties that can also be predicted by scaling from the Sun.

### 4.1. The primary frequency splitting

When the Sun is observed in integrated light, the resulting power spectrum has a regular series of peaks modulated by a broad envelope (see Fig. 4). Each peak in the frequency spectrum corresponds to a harmonic mode characterized by a radial order $n$ and an angular degree $l$. Observations made in light integrated over the stellar disk have highest sensitivity to the low-degree modes ($l = 0, 1$). As shown in Fig. 4, this results in a power spectrum with two superimposed sets of peaks. Theoretical calculations using asymptotic theory (see, e.g., Christensen-Dalsgaard 1988) predict the frequency spectrum to be $\nu_{n,l} \simeq (n + l/2 + \epsilon)\Delta\nu_0$. Here, $\epsilon$ is a constant ($\sim 1.6$ for the Sun) and

$$\Delta\nu_0 \simeq \left(2\int_0^R \frac{dr}{c_s}\right)^{-1}$$

is the primary frequency splitting, approximately equal to the inverse of the sound travel time directly through the star. For the Sun, $\Delta\nu_0$ has a value of $134.92\,\mu\mathrm{Hz}$ (Toutain & Fröhlich 1992).

The adiabatic sound speed satisfies $c_s^2 \propto T$ (Eq. 2), so that $\Delta\nu_0 \propto \sqrt{\langle T\rangle}/R$, where $\langle T\rangle$ is the average internal temperature. Simple estimates for internal values of the pressure and the temperature (e.g., Kippenhahn & Weigert 1990) give $\langle P \rangle \propto M^2/R^4$ and $\langle T \rangle \propto M/R$. We then have

$$\Delta\nu_0 \propto \left(M/R^3\right)^{1/2}.$$

That is, the primary frequency splitting is directly proportional to the mean density of the star. This relationship is well known and gives very good agreement with detailed model calculations (Ulrich 1986) and we use it to scale from the solar value to other stars:

$$\Delta\nu_0 = (M/M_\odot)^{1/2}(R/R_\odot)^{-3/2}\ 134.9\ \mu\text{Hz}. \tag{9}$$

### 4.2. The frequency for maximum power

The power spectrum of solar oscillations is modulated by a broad envelope whose maximum is at a frequency of $\nu_{max} \simeq 3$ mHz (period 5 minutes). The shape of the envelope and the value of $\nu_{max}$ are determined by the excitation and damping. Note that there is a fundamental maximum frequency for oscillations set by acoustic cutoff. Since the acoustic cutoff frequency ($\nu_{ac}$) also defines a typical dynamical timescale for the atmosphere, it has been argued that $\nu_{max}$ in other stars should scale with $\nu_{ac}$ (e.g., Brown et al. 1991). That is, we expect $\nu_{max} \propto c_s/H_P$, where $c_s$ is the sound speed and $H_P \propto T/g$ is the pressure scale height of the atmosphere (Lamb 1932). Under the same assumptions that we used in Sect. 2, we then have $\nu_{max} \propto g/\sqrt{T_{eff}}$. In this way, we can scale from the solar case to predict the frequency of maximum power for an arbitrary star:

$$\nu_{max} = \frac{M/M_\odot}{(R/R_\odot)^2\sqrt{T_{eff}/5777\ K}}\ 3.05\ \text{mHz}. \tag{10}$$

Finally, since we know that the maximum power in the Sun is seen for modes where $n \simeq 21$, we can combine the above equations to predict the $n$-value for maximum power:

$$n_{max} \simeq \left(\frac{M/M_\odot}{(T_{eff}/5777\ \text{K})(R/R_\odot)}\right)^{1/2} \times 22.6 - 1.6. \tag{11}$$

## 5. Observations of solar-like oscillations

There have been many attempts to detect solar-like oscillations in other stars. Table 3 lists parameters of six stars for which measurements have been reported. Some of the observations provided upper limits, but claims for detections have also been made. Table 4 lists the observational results, together with predictions of oscillation amplitudes and frequencies calculated using Eqs.(7)–(11). For several observations, we believe the published amplitudes should be multiplied by the factor shown in Table 4 (see Appendix A.1 for a discussion of amplitude calibration). Data for twelve stars in the open cluster M67 are discussed separately in Sect. 5.7 below.

We now discuss each star in turn, concentrating on observations with the lowest noise levels (i.e., those giving good upper limits or possible detections). We argue that none of the reported detections provides good evidence for oscillations.

Particular attention is given to recent claims for detections on $\alpha$ Cen A (Pottasch et al. 1992) and on Procyon (Brown et al. 1991). This may give the impression that we have singled out a few results for special criticism. We wish to stress that we have treated these papers in detail only because we consider them to be the most important. Several of the other, far less convincing results have already been discussed in the literature.

### 5.1. $\alpha$ Cen A

The $\alpha$ Cen system is the nearest to the Sun and the primary has spectral type G2 V, making it an obvious candidate for solar-like oscillations. Detailed models by Edmonds et al. (1992) predict a frequency splitting of $\Delta\nu_0 = 107.9\ \mu$Hz, in agreement with that obtained from Eq.(9). This star is somewhat more luminous than the Sun and the predicted oscillation amplitudes are therefore $\sim 33\%$ greater than solar (Table 4).

So far, all searches for oscillations in $\alpha$ Cen A have been made using velocity measurements. In no cases have noise levels been achieved that approach the predicted amplitude of $\sim 30\ \text{cm s}^{-1}$. The strongest upper limits have been obtained by Brown & Gilliland (1990) and Edmonds (1993). An earlier positive detection by Gelly et al. (1986) is completely inconsistent with these upper limits, as discussed more fully by Brown & Gilliland (1990).

The best case for oscillations in $\alpha$ Cen A has been made by Pottasch et al. (1992). They found velocity amplitudes 3–5 times greater than solar, which is surprising given the similarity of this star to the Sun. We shall now discuss this paper in some detail. The data were obtained on six consecutive nights: 2–7 April 1990. As part of the analysis they have subdivided the time series into three two-night intervals. Figure 3 of Pottasch et al. shows their CLEANed power spectra for the three subintervals and also for the full six nights. As they point out, these power spectra show many strong peaks and evidence for the periodic comb-like signature that one expects from solar-like oscillations. We first discuss the amplitudes of the peaks, and then consider the possible oscillation signature.

The first point to note is that, due to variations in seeing and wind buffeting of the telescope, the data from the first two nights (2–3 April) have substantially lower noise than from the last four. However, the strongest peaks in the power spectra occur for the second and third pairs of nights (see Figs. 3b and 3c of Pottasch et al.). Thus the signal, if real, appears to know the noise level of the observations. This point is not yet conclusive because we expect signal peaks to be enhanced by noise. We discuss this effect in Appendix A.2 and show that, even when this is taken into account, there is still a strong correlation between the observed signal and the noise. Conversely, we show that the amplitudes of the power spectrum peaks are consistent with the expected noise level.

There is a further problem with identifying so many peaks as oscillation modes, as is proposed by Pottasch et al. The problem is that the frequency range containing the signal (2.25–3.45 mHz) does not contain excess power relative to the rest of the power spectrum. This means that, if most of peaks in this range really come from oscillations, this region of the power spectrum must have a much lower noise level than its surroundings. In other words, there is not enough power to account for both the noise and the signal.

**Table 3.** Stars that have been examined for solar-like oscillations

| Name | $T_{eff}$/K | $L/L_\odot$ | $R/R_\odot$ | $M/M_\odot$ | |
|------|------------|-------------|-------------|-------------|---|
| Sun | $5777 \pm 2.5^a$ | 1.0 | 1.0 | 1.0 | |
| $\alpha$ Cen A | $5770 \pm 20^b$ | $1.45 \pm 0.03^c$ | $1.21 \pm 0.02$ | $1.09 \pm 0.01^c$ | |
| $\beta$ Hyi | $5800 \pm 100^d$ | $2.7 \pm 0.2^d$ | $1.63 \pm 0.08$ | $0.99 \pm 0.05^d$ | |
| Procyon | $6500 \pm 100^e$ | $7.1 \pm 1.0^e$ | $2.10 \pm 0.16$ | $1.50 \pm 0.05^e$  or  $1.75 \pm 0.05^f$ | |
| HD 155543 | $6700 \pm 200^g$ | $3.7 \pm 1.0^g$ | $1.43 \pm 0.21$ | $1.35 \pm 0.05^g$ | |
| $\varepsilon$ Eri | $5180 \pm 50^h$ | $0.33 \pm 0.03^h$ | $0.71 \pm 0.04$ | $0.85 \pm 0.05^h$ | |
| Arcturus | $4300 \pm 30^i$ | $184 \pm 22$ | $24.5 \pm 1.4^i$ | $0.7 \pm 0.3^j$ | |

[a] Stix (1989); [b] Soderblom (1986); [c] Demarque et al. (1986); [d] Dravins et al. (1993);
[e] Guenther & Demarque (1993); [f] Irwin et al. (1992); [g] Belmonte et al. (1990b);
[h] Drake & Smith (1993); [i] Peterson et al. (1993); [j] Bonnell & Bell (1993)

**Table 4.** Predictions and measurements of solar-like oscillations

| Reference | $v_{osc}$ (cm s$^{-1}$) | $(\delta L/L)_{550}$ (ppm) | $(\delta L/L)_{440}$ (ppm) | $\nu_{max}$ (mHz) | $\Delta\nu_0$ ($\mu$Hz) | $n_{max}$ |
|-----------|------------------------|---------------------------|---------------------------|-------------------|------------------------|-----------|
| • Sun (G2 V) | | | | | | |
| see text | $23.4 \pm 1.4$ | $4.7 \pm 0.3$ | $5.9 \pm 0.4$ | 3.05 | 134.9 | 21 |
| • $\alpha$ Cen A (G2 V) | | | | | | |
| prediction | $31.1 \pm 2.0$ | $6.3 \pm 0.4$ | $7.8 \pm 0.5$ | 2.3 | $105.8 \pm 2.7$ | 20 |
| Gelly et al. (1986) | 150 [×2] | — | — | 3.4 | 165.5 | 19 |
| Brown & Gilliland (1990) | < 70–80 | — | — | — | — | — |
| Pottasch et al. (1992) | 75–120 | — | — | 2.9 | 110.6 | 24 |
| Edmonds (1993) | < 50–60 | — | — | — | — | — |
| • $\beta$ Hyi (G2 IV) | | | | | | |
| prediction | $64 \pm 7$ | $12.7 \pm 1.5$ | $15.9 \pm 1.8$ | 1.1 | $64 \pm 5$ | 16 |
| Frandsen (1987) | — | < 50 (Mg I) | — | — | — | — |
| Edmonds (1993) | < 150–200 | — | — | — | — | — |
| • Procyon (F5 IV) | | | | | | |
| prediction (1.50 $M_\odot$) | $111 \pm 17$ | $17.6 \pm 2.8$ | $22.0 \pm 3.5$ | 1.0 | $54 \pm 6$ | 16 |
| prediction (1.75 $M_\odot$) | $95 \pm 15$ | $15.1 \pm 2.4$ | $18.8 \pm 3.0$ | 1.1 | $59 \pm 6$ | 18 |
| Gelly et al. (1986) | 70 [×2] | — | — | 1.2 | 79.4 | 15 |
| Libbrecht (1988) | < 100 | — | — | — | — | — |
| Innis et al. (1991) | < 400 [×$\sqrt{2}$] | — | — | — | — | — |
| Brown et al. (1991) | 50–60 | — | — | 0.85 | (71?) | (10?) |
| Bedford et al. (1993) | 300–1000 | — | — | 0.95 | 70.6 | 12 |
| • HD 155543 (F2 V) | | | | | | |
| prediction | $64 \pm 18$ | $9.6 \pm 2.7$ | $12.0 \pm 3.4$ | 1.9 | $92 \pm 20$ | 19 |
| Belmonte et al. (1990b) | — | 60 | — | 2.0–2.8 | 97.6 | 23 |
| • $\varepsilon$ Eri (K2 V) | | | | | | |
| prediction | $9.1 \pm 1.1$ | $2.3 \pm 0.3$ | $2.8 \pm 0.4$ | 5.3 | $204 \pm 18$ | 24 |
| Noyes et al. (1984) | — | 100×solar (Ca II H & K) | | 1.7 | 172 | 8.4 |
| • Arcturus (K2 III) | | | | | | |
| prediction | $6200 \pm 2800$ | $2200 \pm 1000$ | $2800 \pm 1300$ | 0.004 | $0.9 \pm 0.2$ | 3 |
| Belmonte et al. (1990a) | 6000 | — | — | 0.0043 | (5.0?) | 0–1 |

We now turn discuss whether the data show a regular series of peaks. To investigate this, Pottasch et al. calculate power spectra of power spectra (PS⊗PS) for each of the three two-night subintervals. For the first subinterval, which also has the lowest noise, they find no significant peaks in the PS⊗PS. For 4–5 April, the PS⊗PS shows four regularly spaced peaks (Fig. 4a of Pottasch et al.). However, inspection of the data shows that these are entirely due to the presence of two strong peaks in the power spectrum, the stronger of which is not even identified as an oscillation mode. These two peaks are separated by 707.1 μHz and the four peaks in PS⊗PS occur at $1/n$ times this frequency, where $n = 10, 11, 12$ and 13. The structure in this PS⊗PS is therefore unrelated to the claimed oscillation spectrum.

The power spectrum of the last two nights (6–7 April) has a remarkably regular series of peaks, and this shows up clearly as a single peak in the PS⊗PS. Such a strong peak in the PS⊗PS would usually be taken as evidence for solar-like oscillations. However, given that the amplitudes of the peaks in the power spectrum are not consistent with oscillations, as discussed above, we must ask whether regularly spaced peaks could also be produced by noise. In fact, this is certainly the case, as shown by several examples in the literature. One good example is seen in Fig. 17 of Gilliland et al. (1991), who show PS⊗PS for four stars in M67. In each case they find a single strong peak in PS⊗PS, but in all cases they conclude that the signal is produced by noise fluctuations. One of these stars (No. 37) has since been observed with far higher sensitivity (see Sect. 5.7 below), confirming that Gilliland et al. (1991) were correct in not claiming a detection.

How might noise cause regular peaks in the power spectrum (and therefore a single peak in the PS⊗PS)? By definition, white noise cannot produce this signature but, as is well known, there are many sources of non-white noise. Any effect that causes a noise structure to repeat at regular intervals, either during the observations or in the reduction process, will generate power at the regularly spaced harmonics of this time interval. This can be seen easily by simulating a time series in which some fraction of the noise satisfies $x(t + \Delta t) = x(t)$. The resulting power spectrum has a series of peaks placed at multiples of $1/\Delta t$ and the PS⊗PS has a single strong peak at this frequency splitting. In the case of α Cen A (6–7 April), the time interval corresponding to the observed peak in the PS⊗PS is 4.58 hours which, suggestively, is very close to half the observing time on these two nights.

In conclusion, we do not believe there is good evidence for solar-like oscillations on α Cen A. The upper limits set by observations to date are still 2–3 times greater than the solar amplitude.

## 5.2. β Hyi

This G2 subgiant has the same mass and temperature as the Sun and a luminosity of 2.7 $L_\odot$. We therefore expect oscillation amplitudes in both velocity and luminosity to be about 2.7 times solar. In addition, the star's extreme declination ($-77°$) offers the chance of long uninterrupted time series. Unfortunately, observations to date set upper limits well above the expected signal (see Table 4).

## 5.3. Procyon

Procyon (α CMi) is an F5 subgiant and the second brightest star of near-solar type (the brightest is α Cen). Note that the astrometrically derived mass (1.75 $M_\odot$) disagrees with the mass required to reproduce Procyon's observed luminosity and temperature (1.50 $M_\odot$). We have given predictions for both masses in Table 4. The values we find for $\Delta\nu_0$ are consistent with those found from model calculations by Guenther & Demarque (1993). The oscillation amplitudes are predicted to be 4–5 times solar.

Brown et al. (1991) present evidence for oscillations in Procyon based on velocity measurements taken on six consecutive nights. The power spectrum shows an excess of power in a broad envelope, just as one expects from solar-like oscillations, and in a frequency range consistent with predictions. Most importantly, the amplitudes they infer are actually below those predicted by our scaling relation. Nevertheless, we dispute their claimed detection on the basis that we believe the observations can be completely explained by noise. Of course, this accounts for the absence of a recognizable $p$-mode spectrum in the data.

The key point is that the time series data were high-pass filtered to remove slow variations, using a filter that goes dangerously close to the claimed oscillation signal. Brown et al. argue that this could not explain the shape of their power spectrum. This might be true if the noise were white (i.e., having a flat power spectrum). However, the presence of a non-white noise source might explain the observed power excess. To investigate this, we have simulated a time series with the same characteristics as their observations but containing only noise. We then applied the same high-pass filter as Brown et al., which involves convolving the time series with a Gaussian having a FWHM of 900 s and subtracting this smoothed time series from the original.

The simulations included two noise distributions: white noise (as comes from photon noise and reduction noise) and $1/f$ noise (as would arise from spectrograph drifts), where the latter was simulated as a random walk (see Kjeldsen & Frandsen 1992). The amplitudes of these two noise sources were adjusted so that the final rms noise of the time series after filtering matched the values in Table 1 of Brown et al. The ratio between the white and non-white noise was then the only free parameter in our simulation.

Following Brown et al., we calculated the power spectrum of the full 6-day series and of various subsets. By using noise levels of 2.65 m s$^{-1}$ per minute of observing time for the white noise and $20\,\mathrm{m\,s^{-1}}$ per hour of observation for the $1/f$ noise, we were able to reproduce all their power spectra extremely well (Figs. 2–5 of Brown et al.). For example, in Fig. 5 we show the result when the time series is reduced as two independent segments of three contiguous nights. This is to be compared with Fig. 5 of Brown et al. The agreement is remarkable, especially considering the simplicity of the noise model. Furthermore, the size of the $1/f$-noise component they invoke is completely consistent with estimates by Brown et al. of spectrograph drift.

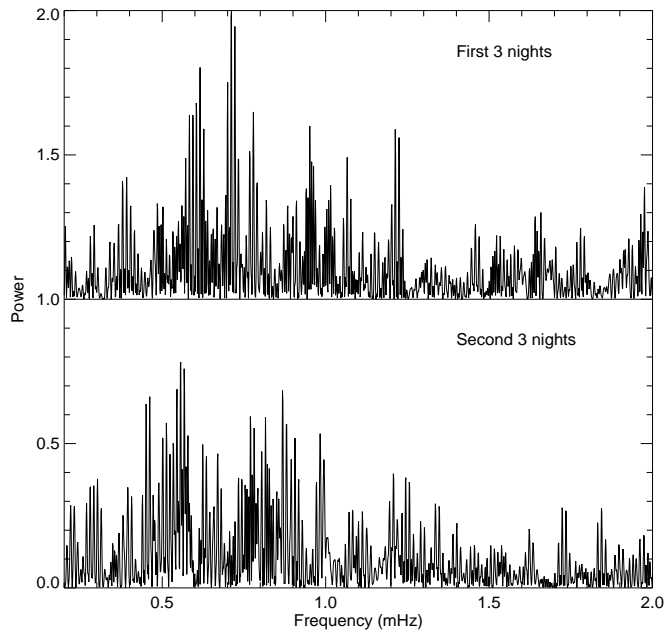

**Fig. 5.** Power spectra of simulated data from Procyon, using a two-component noise model and no signal.

To be fair, Brown et al. did warn that the observed shape of the power spectrum was not a conclusive argument in favour of oscillations, since an unknown noise source could produce a power spectrum with any shape whatsoever. We believe that this is the case and that the power excess in the Procyon data does not provide evidence for stellar oscillations.

One point raised by Brown et al. remains to be explained, namely, why there is no similar excess in the power spectra of the Sun or Arcturus (Figs. 2 and 3 of Brown et al.). This may be due to instrumental effects, since the three sets of data are very different. Alternatively, the non-white noise in the Procyon data may come, at least in part, from granulation noise on the star. In this case, the velocity fluctuations would be of stellar origin but would not arise from coherent oscillations.

If we accept that oscillations on Procyon have not been detected, we are left with a very strong upper limit. The observations and reduction by Brown et al. were thorough and careful and it seems certain that they would have seen oscillations with amplitudes of $50$–$70\,\mathrm{cm\,s^{-1}}$, significantly lower than the predictions. We return to this point in Sect. 6.

Finally, we comment briefly on an earlier claimed detection by Gelly et al. (1986). Firstly, we believe their amplitudes should be multiplied by at least two (their calibration method is not clear from the paper). This means their detection sensitivity is much lower than that of Brown et al. Secondly, their data were also high-pass filtered (the exact details are not given), so it is likely that their power spectrum can also be explained using the above description.

### 5.4. HD 155543

Oscillations in this star were reported by Belmonte et al. (1990b). However, the amplitude is $\sim 6$ times higher than predicted. Note that they found no recognizable signal in the amplitude spectrum: the detection is based on a peak in the power spectrum of the amplitude spectrum. For further discussion on this point, see Gilliland et al. (1991).

### 5.5. ε Eri

Noyes et al. (1984) reported oscillations in this star, but the signal had an amplitude several hundred times higher than expected and was only seen on two out of four nights. The oscillation frequency is also inconsistent with theory. They found no recognizable $p$-mode signature; their value of $\Delta\nu_0$ was based on only a few peaks in the power spectrum.

### 5.6. Arcturus

Arcturus ($\alpha$ Boo) is the brightest northern hemisphere star. It is a red giant, so any oscillations should have much larger amplitudes and longer periods than found in the Sun (Table 4).

Several claims for periodic velocity variations have been made. The strongest, by Belmonte et al. (1990a), is based on a total of 82 hours' data spread over 11 nights. They found evidence for oscillations at several frequencies around a few $\mu$Hz, with the highest peak being at $4.3\,\mu$Hz and having an amplitude of $60\,\mathrm{m\,s^{-1}}$. Both these numbers are in good agreement with the predictions but we feel that, while the periodicity may be of stellar origin, the presence of solar-like oscillations has certainly not been established.

The observed amplitude spectrum shows a $1/f$ structure: the amplitudes of the peaks (Fig. 9 and Table 4 of Belmonte et al.) are well fitted by a $1/f$ noise source having strength $\sim 83/\nu\,\mathrm{m\,s^{-1}}\,\mu\mathrm{Hz}^{-1}$. The authors list several instrumental effects, such as changes in temperature, that could easily be the source of this non-white noise. Furthermore, the same noise structure is seen, with the same amplitude, in the spectrum of solar data taken during daytime (Fig. 7 of Belmonte et al.; note that the frequency axis should be labelled in mHz). We have generated simulated data using a two-component noise source in the same way as described in Sect. 5.3 above. In this case, we obtained good agreement with the observed amplitude spectrum by using a white-noise source having an average amplitude of $30\,\mathrm{m\,s^{-1}}$ per minute of observing time (consistent with Fig. 12b of Belmonte et al.) and a $1/f$ noise of $10\,\mathrm{m\,s^{-1}}$ per hour of observation. An example is shown in the upper panel of Fig. 6. We conclude that the observed amplitude spectrum is well explained by non-white noise.

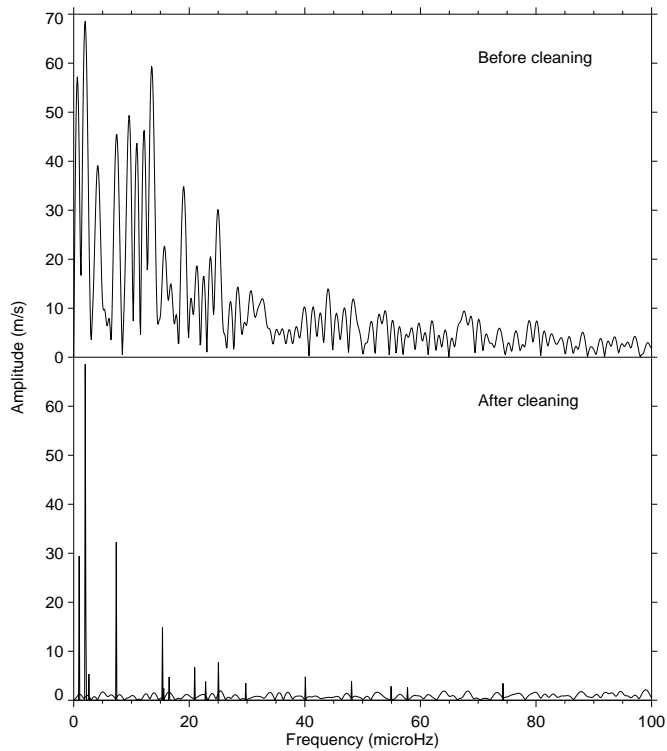

**Fig. 6.** Amplitude spectra of simulated data from Arcturus, using a two-component noise model and no signal.

**Table 6.** Upper limits and predictions for stars in M67

| Star No. | Upper limit (ppm) | Prediction (ppm) | Ratio |
|---|---|---|---|
| 13 | 16 | 22.9 | 0.70 |
| 16 | 18 | 29.8: | 0.60 |
| 27 | 18 | 18.2 | 1.01 |
| 28 | 21 | 13.0 | 1.6 |
| 37 | 21 | 21.8 | 0.95 |
| 41 | 20 | 17.8 | 1.10 |
| 44 | 21 | 13.6 | 1.5 |
| 48 | 23 | 13.3 | 1.7 |
| 49 | 24 | 13.0 | 1.9 |
| 52 | 24 | 12.7 | 1.9 |
| 65 | 37 | 7.7 | 4.8 |

In fact, the observations are inconsistent with a peak amplitude of $60 \, \mathrm{m \, s^{-1}}$ unless there are only one or two excited modes. In other words, while the observations of Belmonte et al. do not give evidence for solar-like oscillations, they do provide a strong challenge to the predictions.

### 5.7. M67

The M67 open cluster was recently the target of a large multi-telescope campaign by Gilliland et al. (1993) using differential CCD photometry. Seven groups collaborated to observe twelve stars over a one-week period. They found suggestive evidence for oscillations on more than half the stars but made no strong claims for detection.

In Table 5 we list properties of the twelve stars, together with our predictions for oscillation amplitudes ($\delta L / L$) and frequencies ($\nu_{\mathrm{max}}$ and $\Delta\nu_0$). We also show the most likely values measured by Gilliland et al. for each of these quantities. These authors give two sets of results based on independent analyses, one using power spectra of CLEANed power spectra (PS$\otimes$PS) and the other using a direct fitting method (comb analysis). In the table we show both sets of results, using three different type styles to indicate the significance attached to each measurement by Gilliland et al.: **bold style** for the most significant and *slanted style* for the least significant. There is almost no agreement between the two analysis methods and, as noted by Gilliland et al., this casts doubt on the reality of the signals.

Note that Gilliland et al. do not claim any unambiguous detections. What upper limits on oscillation amplitudes can be set by the observations? As we discuss in Appendix A.2, a signal with amplitude equal to 2.5 times the noise level would have been seen in the data. Taking this as an upper limit, we obtain the values shown in Table 6, where we have taken the noise levels from Table 11 of Gilliland et al. Note that star no. 12, being a red giant, is a special case and we exclude it from this discussion.

It is clear that many of the suggested amplitudes listed in Table 5 are higher than these $2.5\sigma$ limits, so that one would have

Even if some of the power in the Arcturus data is due to stellar oscillations, the length of the time series is far too short to resolve a solar-like $p$-mode spectrum. The period of the strongest peak is 2.7 days, which is a significant fraction of the total observing time. Scaling to the Sun, this corresponds to observing the five-minute solar oscillations for a total of 6.9 min spread over 22 min. Nevertheless, the authors attempt to identify oscillation modes in the data by cleaning the amplitude spectrum. They find 17 peaks in the range 0–60 $\mu$Hz and identify most of them as oscillation modes. However, the total observing time is 82.3 hours (3.4 $\mu$Hz) and so the frequency interval being examined contains only 18 (= 60/3.4) independent resolution elements. In other words, the cleaning process has simply moved all the $1/f$ power into discrete frequencies. We have found that cleaning our simulated noise spectrum produces peaks with a distribution similar to that seen for the real observations (lower panel of Fig. 6). This confirms that the series of peaks identified by Belmonte et al. is exactly what one would obtain from cleaning $1/f$ noise and gives no evidence for a solar-like oscillation spectrum.

Although the time series is too short allow detection of a $p$-mode spectrum, some of the power at low frequencies may still be due to unresolved stellar oscillations. What upper limit is set by the observations and is it consistent with the predicted amplitude of ($60 \pm 30$) $\mathrm{m \, s^{-1}}$? Our simulations and calculations show that, if Arcturus has a solar-like oscillation spectrum with 10–15 modes, the peak amplitude cannot be more than 30 $\mathrm{m \, s^{-1}}$.

**Table 5.** Parameters and oscillation properties for stars in M67

| Star | $T_{\text{eff}}$ | $L$ | $M$ | $R$ | $(\delta L/L)_{475}$ (ppm) | | | $\nu_{\text{max}}$ (mHz) | | | $\Delta\nu_0$ ($\mu$Hz) | | |
|------|------|------|------|------|------|------|------|------|------|------|------|------|------|
| No. | (K) | ($L_\odot$) | ($M_\odot$) | ($R_\odot$) | pred. | PS⊗PS | comb | pred. | PS⊗PS | comb | pred. | PS⊗PS | comb |
| 12 | 4550 | 19.2 | 1.360 | 7.05 | 124 | **200** | 80 | 0.09 | **0.09** | 0.09 | 8.4 | **13.7** | 7.8 |
| 13 | 6520 | 8.6 | 1.600 | 2.30 | 22.9 | | **16** | 0.87 | *1.50* | **0.39** | 48.9 | *80* | **54.4** |
| 16 | 6050 | 7.7 | 1.280: | 2.53 | 29.8: | **18.5** | *12* | 0.60: | **1.10** | *0.98* | 38.0: | **38.3** | *24.2* |
| 27 | 6120 | 4.8 | 1.280 | 1.95 | 18.2 | **21.5** | 14 | 1.00 | **1.15** | 1.02 | 56.1 | **57.9** | 59.6 |
| 28 | 6520 | 4.2 | 1.375 | 1.61 | 13.0 | **21.5** | 18 | 1.53 | **1.00** | **1.70** | 77.7 | **79.6** | **99.5** |
| 37 | 5170 | 4.3 | 1.340 | 2.59 | 21.8 | **24.5** | 16 | 0.65 | **0.85** | 0.47 | 37.6 | **46.0** | 27.8 |
| 41 | 6120 | 4.7 | 1.280 | 1.93 | 17.8 | **21.5** | 18 | 1.02 | **1.25** | **1.13** | 57.0 | **61.3** | **61.4** |
| 44 | 6160 | 3.6 | 1.265 | 1.67 | 13.6 | | 16 | 1.34 | 1.35 | **1.44** | 70.5 | 62.6 | **58.4** |
| 48 | 6085 | 3.4 | 1.255 | 1.66 | 13.3 | | **16** | 1.35 | 1.35 | **1.23** | 70.7 | 66.7 | **98.5** |
| 49 | 6085 | 3.3 | 1.245 | 1.64 | 13.0 | | *18* | 1.38 | *1.30* | *1.14* | 72.0 | *74* | *71.1* |
| 52 | 6085 | 3.2 | 1.235 | 1.61 | 12.7 | | | 1.42 | *1.70* | | 73.4 | *84* | |
| 65 | 6010 | 1.7 | 1.105 | 1.20 | 7.7 | | | 2.28 | *2.40* | | 107.5 | *117* | |

expected to see the signal directly in the power spectra. Such is the not the case, which casts further doubt on the reality of the oscillations. The fact that genuine signal should be obvious in the power spectrum before it is seen in the PS⊗PS has been noted previously (e.g., Edmonds 1993).

In Table 6 we also show the predicted amplitude for each star and the ratio between this and the observational upper limit. For several stars, the observations rule out oscillations at the predicted level. Further analysis of the M67 data is underway (Kjeldsen et al., in preparation) which should improve these upper limits still further or perhaps detect genuine signals at lower levels. Until then, we conclude that the observations of M67 give important upper limits that, for several stars, are below the predicted amplitude.

## 6. Conclusions

The main results of this paper are summarized in Eqs.(5) and (7). Equation (5), which is based on simple physical arguments and observational data, relates velocity and luminosity amplitudes for many classes of oscillating stars. This relation is important for interpreting photometric observations, such as the recent study of M67, and for deciding whether a given star is best observed in luminosity or velocity. Equation (7) is a fit to the models of Christensen-Dalsgaard & Frandsen (1983) and allows us to scale oscillation amplitudes from the Sun to other stars.

In the Appendix we discuss several factors that must be considered when estimating oscillation amplitudes from observations. One important point is the tendency for noise to reinforce signal peaks in the amplitude spectrum, which might lead one to overestimate the true amplitude (Appendix A.2). In Appendix A.3.1 we examine two effects intrinsic to the star: the finite lifetimes of oscillation modes and the beating between modes closely spaced in frequency. These effects have been invoked by some authors to explain large day-to-day variations in their measurements. For the first effect, we show that dividing a time series into shorter subintervals will not increase the amplitude of a real stellar signal and will generally result in a poorer detection sensitivity. For the second effect, we argue that mode beating will only be significant in certain situations and, in any case, the amplitude modulation will be systematic. Neither effect should be invoked to explain why an observed oscillation mode appears strongly on one night and is not seen again.

How do the amplitudes predicted by Eqs.(5) and (7) compare with observations? We argue that, despite several claimed detections, there is no good evidence for solar-like oscillations on any star except the Sun. For stars similar to the Sun ($\alpha$ Cen A and $\beta$ Hyi), we expect the amplitude predictions to be accurate. Unfortunately, observations of these stars have not reached sufficient sensitivity to detect the amplitudes we predict.

Extrapolation to stars very different from the Sun is less certain. Red giants represent an extreme case. The predicted oscillation amplitudes are large but the periods are long, demanding many months of observing time to obtain resolution sufficient to detect a $p$-mode oscillation spectrum. The most recent searches for solar-like oscillations have concentrated on stars hotter than the Sun, on the basis that their amplitudes should be several times greater than solar. For several F-type stars, namely Procyon and members of M67, detection sensitivities below the predicted amplitudes have now been achieved. The negative observational results indicate that these stars must oscillate with amplitudes less than generally thought. Indeed, Christensen-Dalsgaard & Frandsen (1983) have already noted that mixing-length theory overestimates the convective flux in relatively hot stars and that predictions of oscillation amplitudes may have to be reduced.

It is well accepted that the successful measurement of oscillation frequencies will place important constraints on stellar model parameters (e.g., Brown et al. 1994). Future observations in the style of the M67 campaign could produce an increase in sensitivity of about a factor of two (Gilliland et al. 1993). In addition, new analysis techniques may yet produce detections from the existing M67 data set (Kjeldsen et al., in preparation). Until a genuine detection is made, better estimates for oscillation amplitudes will only come from improved models of the convection process.

*Acknowledgements.* We are very grateful to Ron Gilliland, Jørgen Christensen-Dalsgaard, Peter Edmonds, Lawrence Cram and Albert Zijlstra for valuable comments on this paper, and to Jaymie Matthews for providing information on roAp stars. We thank Signe Kjeldsen for stimulating the discussions. This research has made use of the Simbad database, operated at CDS, Strasbourg, France.

## A. Appendix: Measuring oscillation amplitudes

### A.1. Units of power spectra

The usual way to estimate oscillation amplitudes from observations is by measuring peaks in the power spectrum. There are two ways of choosing the units for the power spectrum. One way is simply to use (amp)$^2$, where 'amp' might be ppm (for luminosity measurements) or m s$^{-1}$ (velocity measurements). We shall refer to this as a 'normal' power spectrum. The other way is to use power density, such as (amp)$^2$/Hz, which we refer to as a power density spectrum (PDS). We stress that the only difference is a change in the way the axis is labelled: the two types of spectra differ only by a multiplicative constant.

When extracting numbers from a plot, it is important to be aware of which system is being used. With a 'normal' power spectrum, estimating amplitudes is straightforward provided the spectrum is properly calibrated (see below): a stellar oscillation with an amplitude of $(\delta L/L) = 10$ ppm will produce a peak of 100 ppm$^2$. In the case of a PDS, inferring the oscillation amplitude from a peak requires that we know the frequency resolution of the observation. If we have a continuous time series of length $T$ seconds, this same 10 ppm oscillation will produce a peak in the PDS of height $100\,T$ ppm$^2$. That is, the strength of the peak in units of power density increases with the observing time.

The formulae given above can be used to infer oscillation amplitudes from a power spectrum, but only if it has been properly calibrated. By this, we mean that a sine wave with amplitude $A$ in the time series produces a peak of strength $A^2$ in the 'normal' power spectrum and a peak of $A^2\,T$ in the PDS. This calibration can be checked by injecting an artificial signal into the time series. The factor of four in the formulae below is a result of doing this calibration correctly (Kjeldsen & Frandsen 1992). Some power spectra in the literature are calculated using Fourier transforms but are not calibrated in the way described. In these cases, deriving correct amplitudes from the published spectra requires that the power levels be multiplied by a factor of two (or sometimes four, depending on the details of the Fourier-transform normalization). Observations for which we believe this correction is necessary are indicated in Table 4.

Finally, there are two useful formulae that describe the noise level in a properly calibrated spectrum (see Kjeldsen & Frandsen 1992). Firstly, the mean noise level in the power spectrum is

$$\sigma_{\mathrm{PS}} = 4\sigma_{\mathrm{rms}}^2/N, \tag{A1}$$

where $N$ is the number of measurements in the time series and $\sigma_{\mathrm{rms}}$ is their rms scatter. Secondly, if the noise is gaussian then the mean noise level in the *amplitude* spectrum (which is the square root of the power spectrum) is:

$$\sigma_{\mathrm{amp}} = \sqrt{\pi\sigma_{\mathrm{PS}}/4}. \tag{A2}$$

### A.2. The effects of noise

In this section we discuss how to estimate oscillation amplitudes in the presence of noise. Suppose we have made observations of a star, calculated the amplitude spectrum (or its square, the power spectrum) and identified several peaks that we believe to be due to stellar oscillations. We now wish to use the strengths of these peaks to calculate $A_{\mathrm{osc}}$, the true oscillation amplitude of the star.

In Fig. 7 we show the amplitude spectra of simulated solar-like oscillations for several different noise levels. The input signal (bottom panel) consists of many oscillating modes modulated by a broad envelope with peak amplitude $A_{\mathrm{osc}} = 1$ unit. In the other panels, we see that many of the signal peaks have been strengthened significantly by constructive interference with noise peaks. Unless we allowed for this effect, we would overestimate the amplitude of the signal.

To quantify the effect, we have performed 100 simulations identical to the one in Fig. 7, with different random number seeds for the noise. For each amplitude spectrum, we measured both the amplitude of the strongest peak ($A_1$) and the average of the five strongest peaks ($A_5$). Consistent with our initial scenario, we included only genuine signal peaks (i.e., peaks that coincided with one of the known oscillation frequencies).

On the basis of these simulations, we can give the following formulae:

$$(A_1)^2 = (A_{\mathrm{osc}})^2 + (8.7 \pm 2.3)\sigma_{\mathrm{amp}}^2 \tag{A3}$$

$$(A_5)^2 = 0.94(A_{\mathrm{osc}})^2 + (3.4 \pm 1.1)\sigma_{\mathrm{amp}}^2, \tag{A4}$$

where $\sigma_{\mathrm{amp}}$ is the mean noise level in the amplitude spectrum. Equations (A3) and (A4) can be used to calculate the 'true' oscillation amplitude ($A_{\mathrm{osc}}$) from the observed amplitude ($A_1$ or $A_5$).

The following points should be noted:

1. The distributions of $A_1$ and $A_5$ for a given noise level are not Gaussian, but have a tail with some high peaks. The scatters given above are rms values.

2. The three dimensionless constants in the equations depend on the choice of the oscillation envelope. We have adopted an envelope similar to that of solar oscillations; a broader envelope would lead to larger values of $A_1$ and $A_5$ because there would be more modes available that could be enhanced by constructive interference with noise peaks.

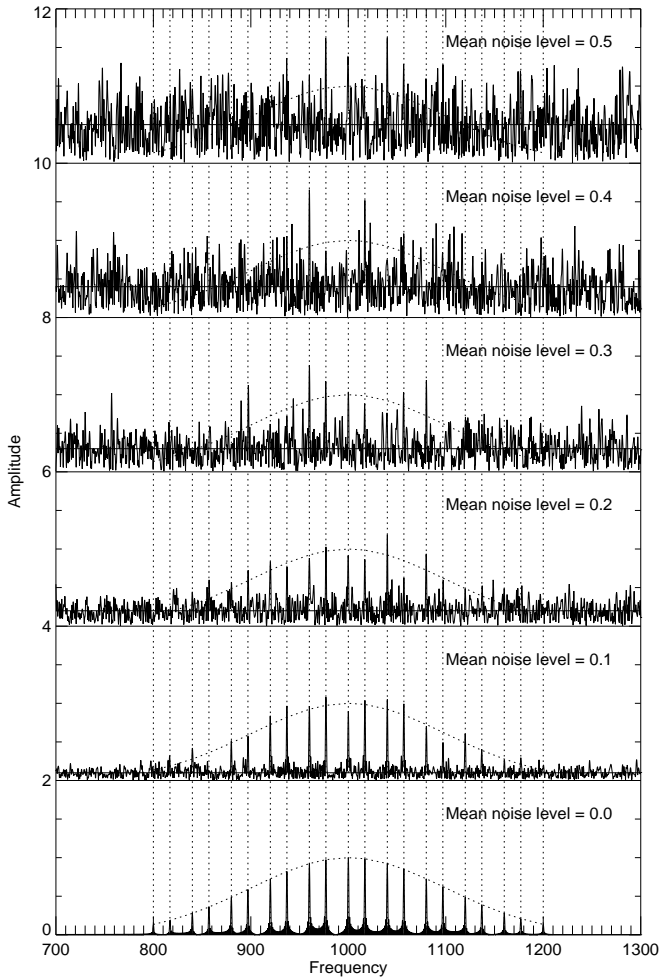

**Fig. 7.** Amplitude spectrum of a solar-like oscillation in the presence of white noise. Dashed lines indicate the frequencies and envelope of the input signal. The mean noise level in each panel is shown by a horizontal line.

3. It might seem that, by increasing the signal, the noise is actually helping us detect the oscillations. This is not the case because, as is evident from the simulations, the signal grows much less quickly than the noise.

From simulations like the ones in Fig. 7, we can say that a signal with amplitude equal to twice the noise level should be seen clearly. Note that this assumes the time series is continuous. In practice this is generally not the case, and both signal and noise peaks will exhibit sidelobes. Additional simulations show that the detectability of the signal is only slightly affected by this and that a signal at a level of 2.5 times the noise is still clearly visible.

### A.2.1. Observations of α Cen A by Pottasch et al. (1992)

As described in Sect. 5.1, the power spectra of these data show many strong peaks that Pottasch et al. identify as oscillation modes. For each of the four power spectra in their Fig. 3, we have calculated $\sigma_{\mathrm{amp}}$ from the rms noise values given in Ta-

**Table 7.** Noise levels and amplitudes for α Cen A (in m s$^{-1}$)

| Dates | $\sigma_{\mathrm{amp}}$ | $A_5$ | $A_{\mathrm{osc}}$ | $A_1$ | | $N_5$ | |
|---|---|---|---|---|---|---|---|
| | | | | E | O | E | O |
| 2–3 April | 0.46 | 1.02 | 0.58 | 1.48 | 1.24 | 1.12 | 1.09 |
| 4–5 April | 0.52 | 1.17 | 0.69 | 1.68 | 1.45 | 1.28 | 1.34 |
| 6–7 April | 0.53 | 1.16 | 0.64 | 1.69 | 1.39 | 1.31 | 1.25 |
| 2–7 April | 0.30 | 0.70 | 0.44 | 0.99 | 0.74 | 0.80 | 0.82 |

E = expected, O = observed

ble 1 of their paper by using Eqs.(A1) and (A2). The results are shown in column (2) of Table 7, while column (3) shows $A_5$, the mean amplitude of the five strongest peaks that are identified by Pottasch et al. as signal. Assuming the signal to be real, we have used Eq.(A4) to estimate its true amplitude. As we see in Table 7, $A_{\mathrm{osc}}$ varies from night to night and, most importantly, the values correlate directly with the noise level. Thus, the signal appears to know the observational noise.

Still assuming the oscillations to be real, we can also use $A_{\mathrm{osc}}$ and Eq.(A3) to calculate the expected value of $A_1$ for each power spectrum. Here, we see that the observed value $A_1$ is lower than expected in all four cases. This results show that identifying the peaks as oscillations is inconsistent with the noise level. The problem is that, even if a genuine signal were present, most of the strongest peaks in the power spectrum should still be noise peaks. We have confirmed this statement using simulations similar to those in Fig. 7.

Given that the peak amplitudes are not consistent with oscillations, are they consistent with noise? For each of the four data subsets, we have generated 50 simulated noise spectra with a mean level of $\sigma_{\mathrm{amp}}$ and then measured $N_5$ for each, where $N_5$ is the mean amplitude of the five strongest peaks. We have also measured $N_5$ for the spectra given by Pottasch et al., this time assuming all peaks to be due to noise. As shown in Table 7, the agreement is good and there are no systematic differences. We conclude that the peaks in the Pottasch et al. data are consistent with a noise level that we calculate directly from the rms scatter in their time series.

### A.3. Amplitude modulation

The previous section describes how noise can alter the observed amplitude of an oscillation. We now discuss sources of amplitude modulation that are intrinsic to the star. Based on observations of the Sun, we can identify two effects that might cause the observed amplitude of a solar-like oscillation to fluctuate from day to day. Firstly, each oscillating mode has a finite lifetime because it is continually being damped and re-excited. Secondly, there is beating between pairs of modes that are closely spaced in frequency.

Both these phenomena have been studied in the Sun using observations (Schrijver et al. 1991; Toutain & Fröhlich 1992) and simulations (Sørensen 1988; Ehgamberdiev et al. 1992; Baudin et al. 1993). Here we discuss an issue not addressed in these papers, namely, whether the detection sensitivity in the presence of amplitude modulation can be improved by subdividing observations into shorter time intervals. For both cases, we argue that the practice of subdividing time series will not improve the S/N and will generally make it worse.

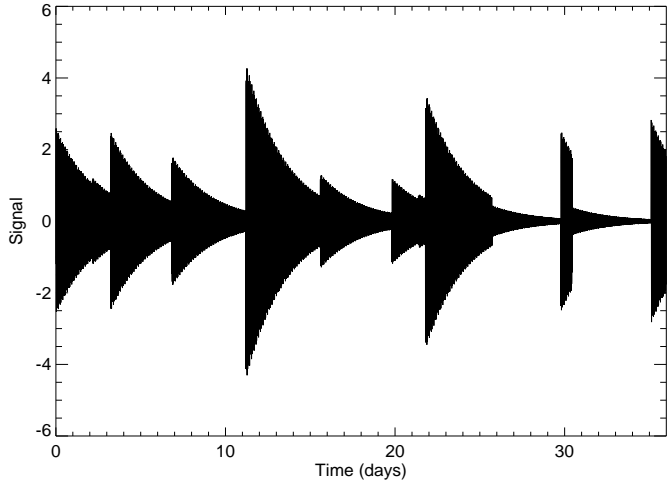

**Fig. 8.** Simulation of a damped oscillation mode being stochastically excited. The damping constant is 3 days. The initial amplitudes of the excitation events are drawn at random from a Gaussian distribution, their initial phases are randomly distributed between 0 to $2\pi$ and the mean interval between excitation events is 3 days.

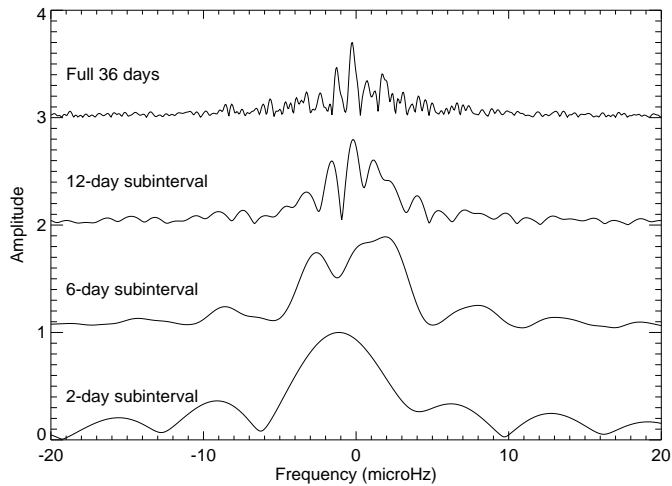

**Fig. 9.** Amplitude spectra of the simulated time series in Fig. 8, calculated using the full 36-day series (top panel) and three subintervals of different lengths.

### A.3.1. Mode lifetimes

Oscillation modes in the Sun have finite lifetimes because they are being continually damped. They are also repeatedly re-excited by convective 'events,' each time with a random phase, with the result that the amplitude of each mode varies on timescales of a few days. We now wish to quantify this effect.

Consider a single oscillation eigenmode that is being continually damped and randomly excited. It is sometimes argued that the signal will be weakened if one observes too long. This would be valid if each new excitation were exactly out of phase with the previous. However, the excitation events have random phases and in this sense, they act in a similar way to noise. Although using a longer time series decreases the total signal (because we have an incoherent sum of many signals with random phases), it also decreases the noise. Thus, there is no reason to expect the S/N to decrease for long time series. Furthermore, noise is completely independent from one subinterval to the next but signal is not (unless the subintervals are very long). We therefore might expect the S/N to increase with the length of observation (up to a limit), despite the fact that the phase of the oscillation is continually changing. Confirmation of this view comes from observations of the Sun: the power spectrum of 160 days of continuous luminosity data (Toutain & Fröhlich 1992) shows very healthy peaks, despite the fact that the lifetimes of the modes are only a few days.

In Fig. 8 we have simulated the oscillation of a single mode as a damped oscillator being forced by a random sequence of excitation events. The details of the simulation are given in the figure caption and are similar to those of Baudin et al. (1993).

Imagine observing this mode, either in velocity or luminosity, with no noise and a continuous 24-hour coverage. The amplitude spectrum for the full 36 days is shown in the top panel of Fig. 9, and we see several distinct peaks modulated by a Lorentz profile. If we take short subintervals of the data, we find fewer peaks in the amplitude spectra. Typical examples for subintervals of $t_{sub} = 12$, 6 and 2 days are also shown in Fig. 9 and are very similar to spectra obtained from observations of the Sun (Toutain & Fröhlich 1992).

From the examples in Fig. 9, we already see that the peak amplitude does not change very much when we change the subinterval length. But these are only single examples and we naturally find a large variation, depending on exactly which subinterval we choose. To quantify this, we have taken six different subinterval lengths ($t_{sub} = 1$, 2, 4, 6, 9 and 18 days) and calculated amplitude spectra for each by moving a window of length $t_{sub}$ through the time series. Each of these amplitude spectra simulates an observation of length $t_{sub}$ centred on a different time. For each one we have measured the peak amplitude; the results are displayed in Fig. 10 as a function of the central time. For $t_{sub} = 36$ days, there is only one 'subinterval' and we have a single point whose height (0.698 units) can be read directly from Fig. 9. For smaller values of $t_{sub}$, we see that the peak amplitude does indeed depend strongly on when we happen to 'observe.'

For each panel in Fig. 10, the dashed line shows the average over all possible subintervals. We have re-plotted these average values in Fig. 11 and see that they decrease rather slowly as $t_{sub}$ increases. By comparison, the amplitude observed from a completely coherent oscillation (i.e., with infinite lifetime) would be independent of $t_{sub}$, as indicated by the horizontal dashed line. On the other hand, white noise decreases in amplitude as $\sqrt{t_{\mathrm{sub}}}$, which is shown for three different noise levels by the sloping dashed lines.

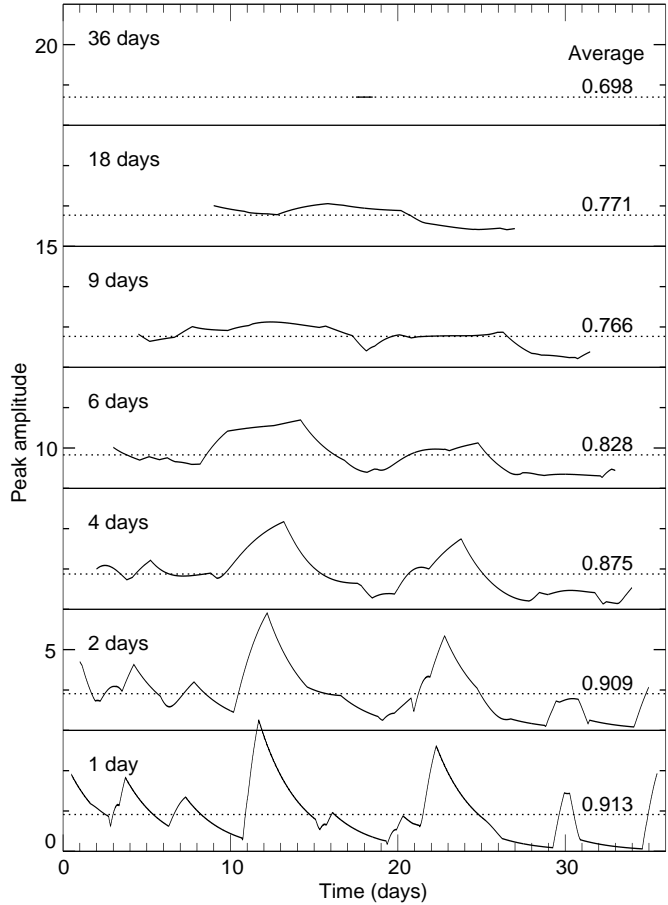

**Fig. 10.** Peak amplitudes obtained from the time series in Fig. 8 when observing windows of different lengths are moved continuously through the data. The average for each case is given at right and is also indicated by a horizontal dashed line.

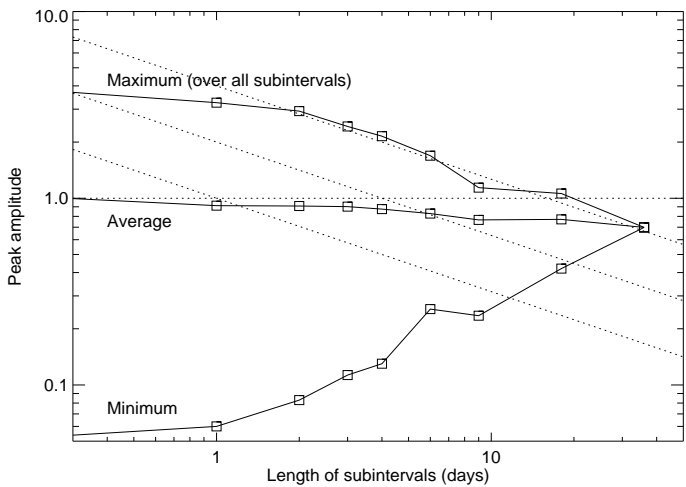

**Fig. 11.** Properties of the peak amplitude that one measures if one subdivided the time series shown in Fig. 8 into intervals of different lengths.

In Fig. 11 we also show the maximum and minimum of each of the curves in Fig. 10, which are the extremes that one might measure from a given subinterval. Thus, the upper points in Fig. 11 (the maximum peak amplitudes) indicate the amplitude one would measure if one happened to choose exactly the 'best' subinterval. Even for this best case, the measured amplitude decreases with $t_{\mathrm{sub}}$ at the same rate as the noise, so that the S/N stays constant. In reality, we would not expect typical measurements to reach this maximum value. Genuine solar-like oscillations will exhibit many independent modes and we expect their amplitudes to be distributed between the limits shown in Fig. 11. In this case the S/N, and hence the detection sensitivity, will always be reduced by subdividing the data.

From the above arguments, we can make the following recommendation for detecting a signal: *one should obtain a time series that is as long as possible and should never expect that subdividing the data will increase the S/N.* Of course, this does not mean one should never subdivide the data for other purposes, but it does mean that the detection sensitivity will not be increased by doing so. We also note that, if one does find a 'signal' that becomes clearer by subdividing the time series, it is probably due to noise (especially if the amplitude of the 'signal' increases as $1/\sqrt{t_{\mathrm{sub}}}$).

Finally, we note that the model of the excitation process of solar oscillations adopted by Sørensen (1988) and Kumar et al. (1988) is different to that used for these simulations. Instead of assuming a sequence of discrete excitation events, as we have done in Fig. 8, these authors model the excitation as a continuous stochastic forcing. We have repeated our simulations using this model – an example of a time series generated in this way is shown in Fig. 12. Using several time series of this type, we have verified that the properties of the power spectra and the conclusions we reached above are unchanged. This is not surprising, given that the present solar data does not allow one to distinguish between these two excitation models.

### A.3.2. Mode beating

To investigate the effect of mode beating, we consider an oscillating star with two modes that have a frequency difference of $10\,\mu$Hz (i.e, beat period 1.16 days). For comparison, the separation between adjacent $l = 0$ and $l = 2$ modes in the Sun is $9\,\mu$Hz (Toutain & Fröhlich 1992). We observe the star continuously with a noiseless system for a time T(obs) and calculate the amplitude spectrum. The result is shown in Fig. 13, where $\Delta\phi$ is the phase difference between the modes at the start of the observations.

When T(obs) is much longer than the beat period (upper panel), the two modes are completely resolved and no amplitude modulation is observed. Even when T(obs) is comparable to the beat period, the beating has very little effect on the observed amplitude. Only when the observing time is less than half a day (bottom two panels) is there significant interference between the modes, and then it is a continuous effect rather than a random fluctuation.

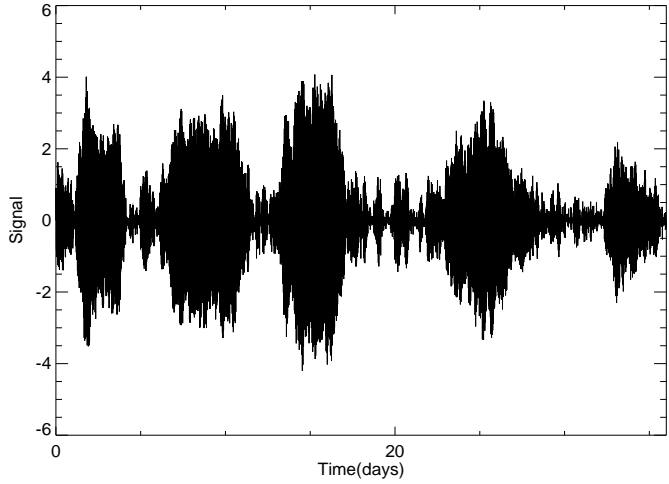

**Fig. 12.** Simulation of a damped oscillation mode being continuously excited. Both amplitude and phase are excited by a Gaussian-distributed force term. The amplitude is continuously damped, with an effective time constant of ∼ 3 days. The total power of the signal shown here is the same as for that in Fig. 8.

Our model contains two modes with equal amplitudes. Data from actual solar-like oscillations will differ in two ways, both of which reduce the importance of mode beating:

1. Expecting amplitudes to be exactly equal is unrealistic because of the stochastic nature of the excitation process (see Sect. A.3.1) and also because $l = 2$ modes are weaker than $l = 0$ modes when observed in integrated light. If the modes have unequal amplitudes, the amplitude modulation caused by beating will be less.
2. The oscillation spectrum will include modes at many different frequencies (Fig. 4). Of these, the $l = 1$ modes may be split by stellar rotation and each $l = 0$ mode will have an $l = 2$ mode close to it. All these modes are excited independently with random phases so that, if we consider the oscillation spectrum as a whole, the effects of beating should average out and the distribution of mode amplitudes should be similar from one observation to the next.

To summarize, we only expect mode beating to be significant for observing times less than the beat period. In this case, amplitudes will change slowly and systematically, but the amplitude distribution (averaged over the oscillation spectrum) will be constant. As in the previous section, we conclude that genuine signal will not become clearer by subdividing the data series and that should be invoked to explain why an observed oscillation mode appears strongly on one night and is not seen again.

### A.4. The solar oscillation amplitude

#### A.4.1. Velocity measurements

The values in Table 1 were derived as follows:

1. Grec et al. (1983): we take the average of the highest peaks in their Fig. 1 and Table 2
2. Isaak et al. (1989): the Na I data in their Fig. 5 and Table 1 indicates an envelope with peak amplitude ∼ 21 cm s⁻¹.
3. Jiménez et al. (1990): the strongest modes ($l = 1$) have an envelope with a peak of ∼ 24 cm s⁻¹ (their Table 1).
4. Libbrecht & Woodard (1991): from their Fig. 8, the peak velocity is 18 cm s⁻¹. This is an rms value and must be multiplied by $\sqrt{2}$.

As discussed in Appendix A.2, noisy data can lead one to overestimate the amplitude of the signal in the power spectrum. This effect explains the higher value of ∼ 35 cm s⁻¹ reported by Pottasch et al. (1992). Their observations were made during an instrumental test and used a strong neutral density filter to simulate the signal expected from $\alpha$ Cen A, resulting in a high noise level in the power spectrum (15–20 cm s⁻¹). Once this is allowed for, the Pottasch et al. value is consistent with the other measurements, which all have noise levels below ∼ 3 cm s⁻¹ and for which this correction is insignificant.

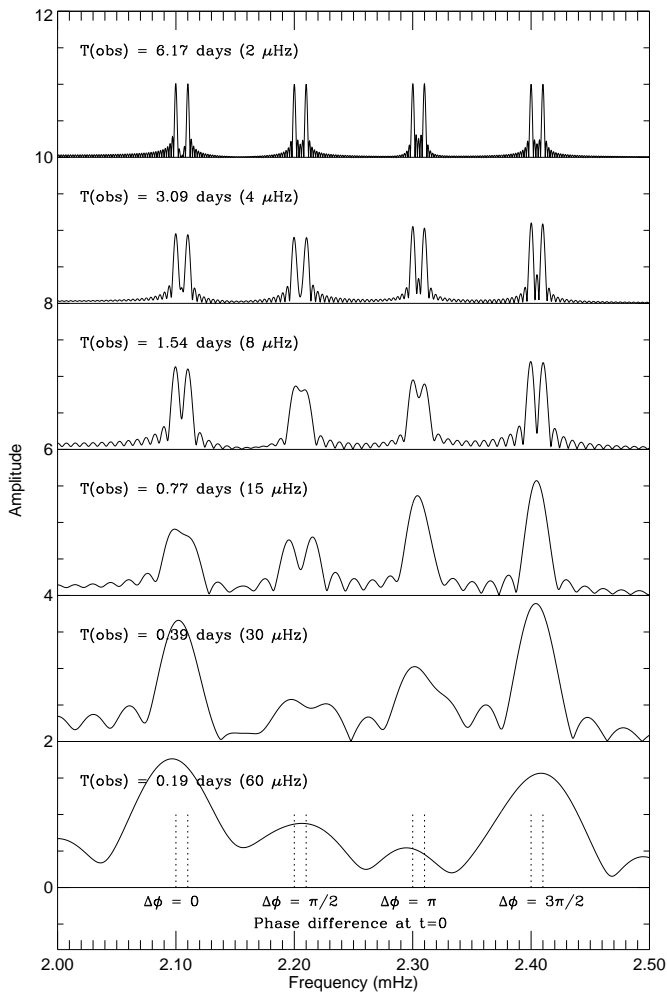

**Fig. 13.** Mode-Beating simulations

A.4.2. Luminosity measurements

The values in Table 2 come from the following observations:

1. Toutain & Fröhlich (1992) have analysed observations made with the IPHIR experiment on board the USSR PHOBOS Mission. Excellent data at 500 nm were obtained over 160 days. At this high resolution, individual modes are completely resolved into distinct peaks modulated by a Lorentz profile (see Appendix A.3.1). To estimate the power that would be observed if the modes were unresolved (i.e., if the time series were shorter), we should integrate under the Lorentz profiles. For the two strongest modes (Fig. 2 of Toutain & Fröhlich), the total rms power for each is 8.1 ppm². We must then take the square root to convert power to amplitude and multiply by $\sqrt{2}$ to convert rms-amplitude to sine-amplitude. We arrive at a peak amplitude of 4.0 ppm. Scaling this to 550 nm (see Sect. 2.2) then gives $(\delta L/L)_{550} = 3.6$ ppm.

2. Woodard & Hudson (1983a,b) have obtained ∼ 10 months of bolometric measurements from the ACRIM instrument on the Solar Maximum Mission satellite. From the $l = 1$ peaks in the frequency range 2.686–3.517 mHz, they found the rms power to be 5.8 ppm². The corresponding sine amplitude is $(\delta L/L)_{bol} = 3.41$ ppm. From their figure it is clear that the highest peaks in this frequency range are 20–25% above the mean, so we adopt $(\delta L/L)_{bol} = 4.2$ ppm as an estimate for the maximum peak. Using Eq.(4) then gives $(\delta L/L)_{550} = 4.7$ ppm.

3. Jiménez et al. (1990) made ground-based observations in three channels (516, 680 and 770 nm). We only consider the 516 nm data, since it has the best S/N. These authors list amplitudes of peaks for $n = 12$–30, with the strongest $(n = 20$–21) having values of 10–11 ppm. However, this overestimates the true signal because of the effect of noise (see Appendix A.2). To correct for this, we need to estimate the average noise level in the amplitude spectrum. From the amplitudes listed by Jiménez et al., especially for peaks with extreme values of $n$ (which have little or no real signal), we estimate $\sigma_{amp} = 4$–5 ppm. Using Eq.(A4) with $A_5 = 10$ ppm and then scaling to 550 nm gives 3.7–6.5 ppm as the amplitude of the underlying oscillation.